\documentclass{article}
\usepackage{spconf,amsmath,graphicx, amsfonts}

\usepackage{caption} 	
\usepackage{multirow, makecell} 	
\usepackage{booktabs} 	
\usepackage{bm}
\usepackage{hyperref}

\def\CLS{\texttt{[CLS]}}

\title{SPEECH RECOGNITION BY SIMPLY FINE-TUNING BERT}
\name{Wen-Chin Huang$^{1,2}$, Chia-Hua Wu$^2$, Shang-Bao Luo$^2$, Kuan-Yu Chen$^3$, Hsin-Min Wang$^2$, Tomoki Toda$^1$}
\address{$^1$Nagoya University, Japan $^2$Academia Sinica, Taiwan\\
		$^3$National Taiwan University of Science and Technology, Taiwan}

\begin{document}
\ninept

\maketitle
\begin{abstract}
We propose a simple method for automatic speech recognition (ASR) by fine-tuning BERT, which is a language model (LM) trained on large-scale unlabeled text data and can generate rich contextual representations. Our assumption is that given a history context sequence, a powerful LM can narrow the range of possible choices and the speech signal can be used as a simple clue. Hence, comparing to conventional ASR systems that train a powerful acoustic model (AM) from scratch, we believe that speech recognition is possible by simply fine-tuning a BERT model. As an initial study, we demonstrate the effectiveness of the proposed idea on the AISHELL dataset and show that stacking a very simple AM on top of BERT can yield reasonable performance. 
\end{abstract}
\section{Introduction}
\label{sec:intro}

Conventional automatic speech recognition (ASR) systems consist of multiple separately optimized modules, including an acoustic model (AM), a language model (LM) and a lexicon. In recent years, end-to-end (E2E) ASR models have attracted much attention, due to the believe that jointly optimizing one single model is beneficial to avoiding not only task-specific engineering but also error propagation. Current main-stream E2E approaches include connectionist temporal classification (CTC) \cite{CTC}, neural transducers \cite{RNN-T}, and attention-based sequence-to-sequence (seq2seq) learning \cite{LAS}.

LMs play an essential role in ASR. Even the E2E models that implicitly integrate LM into optimization can benefit from LM fusion. It is therefore worthwhile thinking: \textit{how can we make use of the full power of LMs?} Let us consider a situation that we are in the middle of transcribing a speech utterance, where we have already correctly recognized a sequence of history words, and we want to determine what the next word is being said. From a probabilistic point of view, a strong LM, can then generate a list of candidates where each of them is highly possible to be the next word. The list may be extremely short that there is only one answer left. As a result, we can use few to no clues in the speech signal to correctly recognize the next word.

There has been rapid development of LMs in the field of natural languages processing (NLP), and one of the most epoch-making approach is BERT \cite{BERT}. Its success comes from a framework where a pretraining stage is followed by a task-specific fine-tuning stage. Thanks to the un-/self-supervised objectives adopted in pretraining, large-scale \textit{unlabeled} datasets can be used for training, thus capable of learning enriched language representations that are powerful on various NLP tasks. BERT and its variants have created a dominant paradigm in NLP in the past year \cite{openai-gpt, roberta, albert}.

In this work, we propose a novel approach to ASR, which is to simply fine-tune a pretrained BERT model. Our method, which we call BERT-ASR, formulates ASR as a classification problem, where the objective is to correctly classify the next word given the acoustic speech signals and the history words. We show that even an AM that simply averages the frame-based acoustic features corresponding to a word can be applied to BERT-ASR to correctly transcribe speech to a certain extent, and the performance can be further boosted by using a more complex model.

\begin{figure*}[t]

\begin{minipage}[b]{0.4\textwidth}
  \centering
  \includegraphics[height=2in]{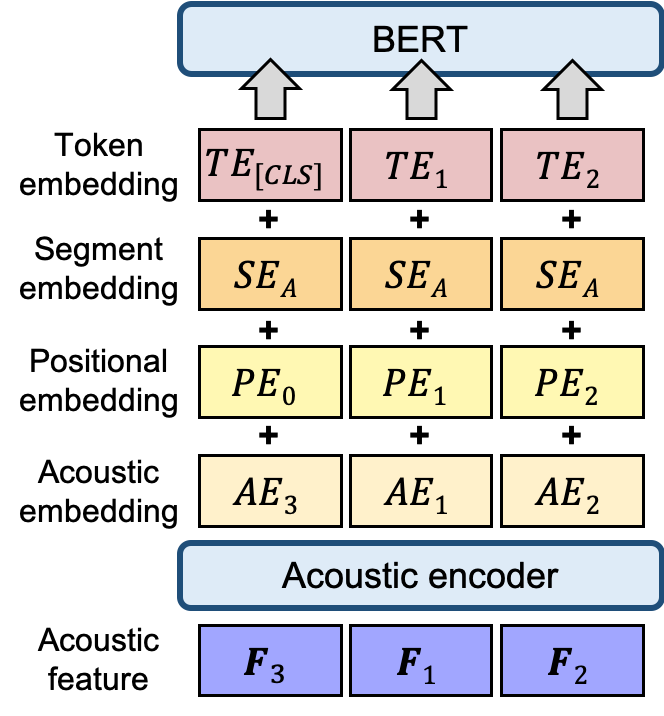}
  \captionof{figure}{The input representation of the original BERT and the proposed BERT-ASR.}
  \label{fig:input-repr}
\end{minipage}
~
\begin{minipage}[b]{0.6\textwidth}
  \centering
  \includegraphics[height=2.2in]{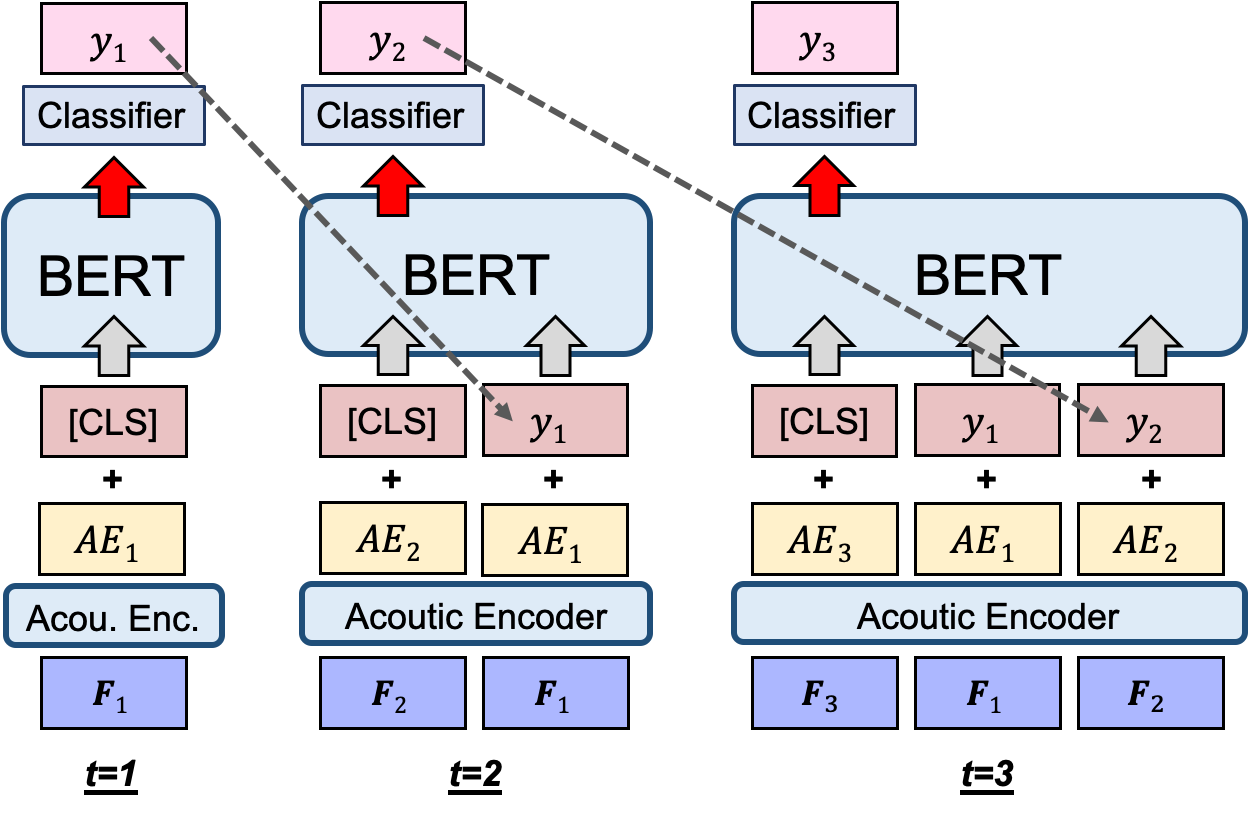}
  \captionof{figure}{Illustration of the decoding process of the proposed BERT-ASR.}
  \label{fig:bert-asr-decoding}
\end{minipage}

\end{figure*}

\section{BERT}
\label{sec:prelim}

BERT \cite{BERT}, which stands for Bidirectional Encoder Representations from Transformers, is a pretraining method from a LM objective with a Transformer encoder architecture \cite{transformer}. The full power of BERT can be released only through a pretraining\textendash fine-tuning framework, where the model is first trained on a large-scale unlabeled text dataset, and then all/some parameters are fine-tuned on a labeled dataset for the downstream task.

The original usage of BERT mainly focused on NLP tasks, ranging from token-level and sequence-level classification tasks, including question answering \cite{k-bert, bert-hae}, document summarization \cite{bert-extractive-summarization, discourse}, information retrieval \cite{twinbert, passage-re-ranking}, machine translation \cite{bert-NMT, bert-score}, just to name a few. There has also been attempts to combine BERT in ASR, including rescoring \cite{bert-asr-rescoring, masked-lm-scoring} or generating soft labels for training \cite{bert-asr-distill}. In this section, we review the fundamentals of BERT.

\subsection{Model architecture and input representations}

BERT adopts a multi-layer Transformer \cite{transformer} encoder, where each layer contains a multi-head self-attention sublayer followed by a positionwise fully connected feedforward network. Each layer is equipped with residual connections and layer normalization.

The input representation of BERT was designed to handle a variety of down-stream tasks, as visualized in Figure~\ref{fig:input-repr}. First, a token embedding is assigned to each token in the vocabulary. Some special tokens were added to the original BERT, including a classification token (\texttt{[CLS]}) that is padded to the beginning of every sequence, where the final hidden state of BERT corresponding to this token is used as the aggregate sequence representation for classification tasks, and a separation token (\texttt{[SEP]}) for separating two sentences. Second, a segment embedding is added to every token to indicate whether it belongs to sentence \texttt{A} or \texttt{B}. Finally, a learned positional embedding is added such that the model can be aware of information about the relative or absolute position of each token. The input representation for every token is then constructed by summing the corresponding token, segment, and position embeddings. 

\subsection{Training and fine-tuning}

Two self-supervised objectives were used for pretraining BERT. The first one is the masked language modeling (MLM), which is a denoising objective that asks the model to reconstruct randomly masked input tokens based on context information. Specifically, 15\% of the input tokens are first chosen. Then, each token is (1) replaced with \texttt {[MASK]} for 80\% of the time, (2) replaced with a random token for 10\% of the time, (3) kept unchanged for 10\% of the time.

During fine-tuning, depending on the downstream task, minimal task-specific parameters are introduced so that fine-tuning can be cheap in terms of data and training efforts.

\section{Proposed Method}

In this section, we explain how we fine-tune a pretrained BERT to formulate LM, and then further extend it to consume acoustic speech signals to achieve ASR.

Assume we have an ASR training dataset containing $N$ speech utterances: $\bm{D}_{\text{ASR}} = \langle \bm{X}^{(i)}, \bm{y}^{(i)}\rangle_{i=1}^N$, with each $\bm{y} = (y_1, ..., y_T)$ being the transcription consisting of $T$ tokens, and each $\bm{X} = (\bm{x}_1, ..., \bm{x}_{T'})$ denoting a sequence of $T'$ input acoustic feature frames. The acoustic features are of dimension $d$, i.e., $\bm{x}_t \in \mathbb{R}^d$, and the tokens are from a vocabulary of size $V$.

\subsection{Training a probabilistic LM with BERT}
\label{ssec:bert-lm}

We first show how we formulate a probabilistic LM using BERT, which we will refer to as BERT-LM. The probability of observing a symbol sequence $\bm{y}$ can be formulated as:
\begin{align}
\label{eq:LM}
P(\bm{y}) = P (y_1, \dots, y_T) = \prod_{t=1}^{T}{P(y_t|y_1,\dots, y_{t-1})}.
\end{align}
The decoding (or scoring) of a given symbol sequence then becomes an iterative process that calculates all the terms in the product, which is illustrated in Figure~\ref{fig:bert-asr-decoding}. At the $t$-th time step, the BERT model takes a sequence of previously decoded symbols and the \texttt{[CLS]} token as input, i.e., $(\texttt{[CLS]}, y_1,\dots,y_{t-1})$.
Then, the final hidden state corresponding to $ \texttt{[CLS]}$ is sent into a linear classifier, which then outputs the probability distribution $P(y_t|y_1,\dots, y_{t-1})$.

To train the model, assume we have a training dataset with $N$ sentences: $\bm{D}_{\text{text}} = \{\bm{y}^{(i)}\}_{i=1}^N$. An essential technique to train the model is to \textit{exhaustively} enumerate all possible training samples. That is, each sentence with $T$ symbols is extended to $T$ different training samples following the rule in Equation~\eqref{eq:cls}:
\begin{align}
\label{eq:cls}
	(y_1, ..., y_T) \to 
		\begin{cases}
			(\texttt{[CLS]})  \\
			(\texttt{[CLS]}, y_1)  \\
			(\texttt{[CLS]}, y_1, y_2)  \\
			\dots  \\
			(\texttt{[CLS]}, y_1,\dots,y_{t-1}) 
		\end{cases}.
\end{align}
The training of the BERT-LM becomes simply minimizing the following cross-entropy objective:
\begin{align}
\label{eq:loss-LM}
\mathcal{L}_{\text{LM}} = - \sum_{i=1}^{N} \sum_{t=1}^{T} P(y^{(i)}_t|\texttt{[CLS]}, y^{(i)}_1,\dots, y^{(i)}_{t-1}).
\end{align}

\begin{figure*}[t]
\begin{minipage}[b]{0.4\textwidth}
  \centering
  \includegraphics[height=1.8in]{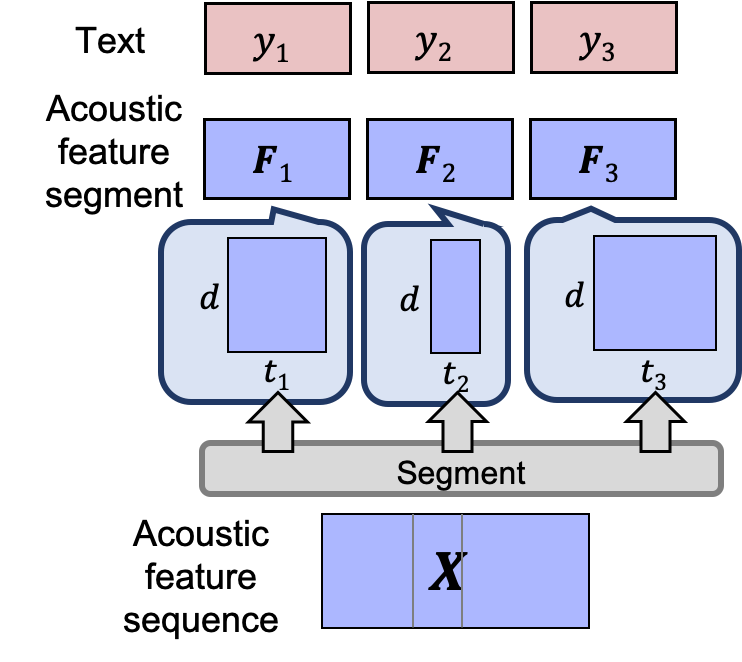}
  \captionof{figure}{Illustration of the alignment between the text and the acoustic frames.}
  \label{fig:alignment}
\end{minipage}
~
\begin{minipage}[b]{0.25\textwidth}
  \centering
  \includegraphics[height=2in]{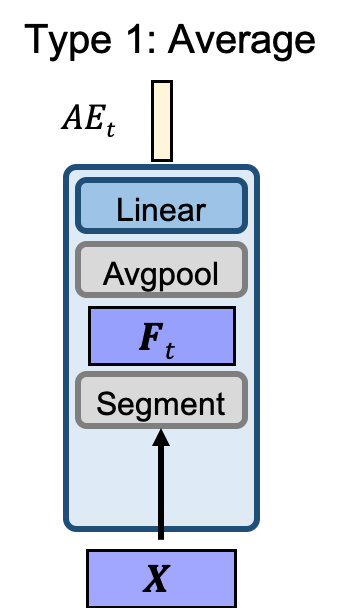}
  \captionof{figure}{The average encoder.}
  \label{fig:avg-enc}
\end{minipage}
~
\begin{minipage}[b]{0.35\textwidth}
  \centering
  \includegraphics[height=2in]{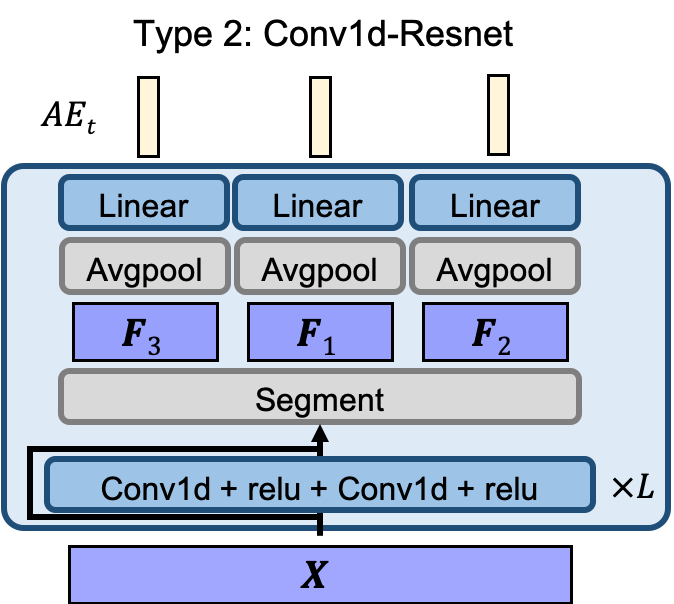}
  \captionof{figure}{The conv1d resnet encoder.}
  \label{fig:conv1d-resnet-enc}
\end{minipage}
\end{figure*}

\subsection{BERT-ASR}
\label{ssec:bert-asr}

We introduce our proposed BERT-ASR in this section. Since the time resolution of text and speech is at completely different scales, for the model described in Section~\ref{ssec:bert-lm} to be capable of taking acoustic features as input, we first assume that we know the nonlinear alignment between an acoustic feature sequence and the corresponding text, as depicted in Figure~\ref{fig:alignment}. Specifically, for a pair of training transcription and acoustic feature sequence $\langle (\bm{x}_1,\dots,\bm{x}_{T'}), (y_1,\dots,y_T)\rangle$, we denote $\bm{F}_i$ to be the segment of features corresponding to $y_i$: $\bm{F}_i=(\bm{x}_{t_{i-1}+1},\dots,\bm{x}_{t_i})\in \mathbb{R}^{t_i \times d}$, which is from frame $t_{i-1}+1$ to $t_i$, and we set $t_0=0$. Thus, the $T'$ acoustic frames can be segmented into $T$ groups: $\bm{X}=(\bm{F}_1,\dots,\bm{F}_T)$, and a new dataset containing segmented acoustic feature and text pairs can be obtained: $\bm{D}_{\text{seg}} = \langle \bm{F}^{(i)}, \bm{y}^{(i)}\rangle_{i=1}^N$.

We can then augment the BERT-LM into BERT-ASR by injecting the acoustic information extracted from $\bm{D}_{\text{seg}}$ into BERT-LM. Specifically, as depicted in Figure~\ref{fig:input-repr}, an acoustic encoder, which will be described later, consumes the raw acoustic feature segments to generate the \textit{acoustic embeddings}. They are summed with the three types of embeddings in the original BERT, and further sent into BERT. Note that the acoustic embedding corresponding to the current word to be transcribed is added to the \CLS\ token as shown in Figure~\ref{fig:bert-asr-decoding}.

The probability of observing a symbol sequence $\bm{y}$ in Equations~\eqref{eq:LM} can therefore be reformulated as:
\begin{align}
\label{eq:ASR}
P(\bm{y}) = \prod_{t=1}^{T}{P(y_t|y_1,\dots,y_{t-1},\bm{F}_1,\dots,\bm{F}_{t})}.
\end{align}
Note that the acoustic segment of the current time step is also taken into consideration, which is essential for the model to correctly transcribe the current word being said. The training objective can be derived by reformulating Equation~\eqref{eq:loss-LM} as:
\begin{align}
\label{eq:loss-ASR}
\mathcal{L}_{\text{ASR}} =& - \sum_{i=1}^{N} \sum_{t=1}^{T} P(y^{(i)}_t| \nonumber \\
&\langle \CLS, \bm{F}_t\rangle, \langle y^{(i)}_1,\bm{F}_1\rangle,\dots, \langle  y^{(i)}_{t-1}, \bm{F}_{t-1}\rangle).
\end{align}

In a nutshell, the training of BERT-ASR involves three steps.
\begin{enumerate}
	\item Pretrain a BERT using a large-scale text dataset.
	\item fine-tune a BERT-LM on the transcriptions of the ASR training dataset, as described in Section~\ref{ssec:bert-lm},
	\item fine-tune a BERT-ASR using both text and speech data of the ASR training dataset.
\end{enumerate}

\subsection{Acoustic encoder}

We now describe two kinds of architectures for the acoustic encoder mentioned in Section~\ref{ssec:bert-asr}. Formally, the acoustic encoder takes the whole acoustic frame sequence $\bm{X}$ as input, and outputs the corresponding acoustic embeddings $(AE_1, \dots, AE_{T})$, where $AE_t\in\mathbb{R}^{d_{model}}$ with $d_{model}$ being the BERT embedding dimension. The acoustic encoder must contain the \textit{segment} layer to obtain the acoustic segments.

\subsubsection{Average encoder}

We first consider a very simple average encoder, as depicted in Figure~\ref{fig:avg-enc}. First, the segmentation is performed.
Then, for each $\bm{F}_t$, we average on the time axis, and then the resulting vector is passed through a linear layer to distill the useful information, while scaling the dimension from $d$ to $d_{model}$.
Simple as it seems, as we will show later, initial results can already be obtained with this average encoder.

\subsubsection{Conv1d resnet encoder}

The drawback of the average encoder is that temporal dependencies between different acoustic segments was not considered. Therefore, we investigate a second encoder, which we will refer to as the conv1d resnet encoder, as illustrated in Figure~\ref{fig:conv1d-resnet-enc}. While it has the identical segment and linear layers as the average encoder,  we add $L$ learnable residual blocks that operates on $\bm{X}$. Each residual block contains two conv1d layers over the time axis followed by ReLU activations. It is expected that taking the temporal relationship between segments into account can boost performance.

\begin{table*}[t]
	\centering
	\captionsetup{justification=centering}
	\caption{Results on the AISHELL-1 dataset. "Orac." and "Prac." denote the oracle decoding and practical decoding, respectively. "Conv1d resnet X" denotes the conv1d resnet encoder with X resnet blocks. Best performance of the BERT-ASR are shown in bold.}
	\centering
	\begin{tabular}{ c c c c c c c c | c c c c }
		\toprule
		\multirow{2}{*}[-2pt]{Model} & \multirowcell{2}[-2pt]{Acoustic encoder} & \multicolumn{2}{c}{Perplexity} & \multicolumn{2}{c}{CER (Orac.)} & \multicolumn{2}{c}{CER (Prac.)} & \multicolumn{2}{c}{SER (Orac.)} & \multicolumn{2}{c}{SER (Prac.)} \\
		\cmidrule(lr){3-4} \cmidrule(lr){5-6} \cmidrule(lr){7-8} \cmidrule(lr){9-10} \cmidrule(lr){11-12}
		 & & Dev & Test & Dev & Test & Dev & Test & Dev & Test & Dev & Test \\
		\midrule
		Trigram-LM & - &  133.32 & 127.88 & \multicolumn{2}{c}{-} & \multicolumn{2}{c}{-} & \multicolumn{2}{c}{-} & \multicolumn{2}{c}{-} \\
		LSTM-LM & - & 79.97 & 78.80 & \multicolumn{2}{c}{-} & \multicolumn{2}{c}{-} & \multicolumn{2}{c}{-} & \multicolumn{2}{c}{-} \\
		BERT-LM & - & 39.74 & 41.72 & \multicolumn{2}{c}{-} & \multicolumn{2}{c}{-} & \multicolumn{2}{c}{-} & \multicolumn{2}{c}{-} \\
		\midrule
		\multirow{5}{*}[-1pt]{BERT-ASR} & Average & 5.88 & 9.02 & 65.8 & 68.9 & 96.4 & 105.8 & 60.3 & 63.5 & 91.5 & 100.3\\
		& Conv1d resnet 1 & 4.91 & 7.63 & 55.8 & 59.0 & 89.6 & 99.6 & 50.0 & 53.8 & 84.4 & 94.1 \\
		& Conv1d resnet 2 & \textbf{4.77} & \textbf{6.94} & \textbf{54.6} & \textbf{58.8} & 89.7 & \textbf{99.1} & 49.5 & 53.6 & 84.6 & 93.5 \\
		& Conv1d resnet 3 & 4.83 & 7.41 & 54.8 & 58.9 & 89.8 & 99.4 & 49.6 & 53.6 & 84.6 & 93.9 \\
		& Conv1d-resnet 4 & 4.78 & 7.29 & \textbf{54.6} & 59.0 & \textbf{89.5} & 99.3 & 49.4 & 53.9 & 84.4 & 93.8 \\
		\midrule
		GMM-HMM & - & \multicolumn{2}{c}{-} & \multicolumn{2}{c}{-} & 10.4 & 12.2 & \multicolumn{2}{c}{-} & \multicolumn{2}{c}{-} \\
		DNN-HMM & - & \multicolumn{2}{c}{-} & \multicolumn{2}{c}{-} & 7.2 & 8.4 & \multicolumn{2}{c}{-} & \multicolumn{2}{c}{-} \\
		\bottomrule
	\end{tabular}
	\label{tab:aishell}
    \vspace*{-3.5mm}
\end{table*}

\section{Experiments}

\subsection{Experimental settings}

We evaluate the proposed method on the AISHELL-1 dataset \cite{aishell-1}, which contains 170 hr Mandarin speech. We used the Kaldi toolkit \cite{kaldi} to extract 80-dim log Mel-filter bank plus 3-dim pitch features and normalized them. The training data contained around 120k utterances, and the exhaustive enumeration process described in Section~\ref{ssec:bert-lm} resulted in 1.7M training samples.
For the first step of the proposed BERT-ASR, i.e., pretraining a BERT model using a large-scale text dataset (cf. Section~\ref{ssec:bert-asr}), we adopt an updated version of BERT, which is whole word masking (WWM), whose effectiveness was verified in \cite{bert-wwm-chinese}. The major difference between the updated BERT and the classic BERT is in the masking procedure of MLM training. If a masked token belongs to a word, then all the tokens that complete the word will be masked altogether. This is a much more challenging task, since the model is forced to recover the whole word rather than just recovering tokens. we directly used the \texttt{hfl/chinese-bert-wwm} pretrained model provided by \cite{bert-wwm-chinese}\footnote{\url{https://github.com/ymcui/Chinese-BERT-wwm}}, which was trained on Chinese Wikipedia. The modeling unit was Mandarin character. We conducted the experiments using the HuggingFace Transformer toolkit \cite{hft}. The alignment we used during training was obtained by forced alignment with an HMM/DNN model trained on the same AISHELL-1 training set.

We considered two decoding scenarios w.r.t the alignment strategy. First, the \textit{oracle decoding} is where we assumed that alignment is accessible.
Second, to match a \textit{practical decoding} setting, as a naive attempt, we assumed that the alignment between each utterance and the underlying text is linear, and partitioned the acoustic frames into segments of equal lengths. The length was calculated as the average number of frames per word in the training set, which was 25 frames. In both scenarios, we considered beam decoding with a beam size of 10. 

\vspace*{-3.5mm}
\subsection{Main results}

We reported the perplexity (PPL) and character error rate (CER) in Table~\ref{tab:aishell}, with the former being a metric to compare the performance of LMs and the latter to compare performance of ASR systems. As a reference, we first compared the PPL between different LMs. It can be clearly observed that the BERT-LM outperformed the conventional trigram-LM and LSTM-LM, again showing the power of BERT as a LM.

We then compare BERT-ASR with BERT-LM. By using a simple average encoder, a significantly lower PPL could be obtained, showing that using acoustic clues can greatly help guide recognition. Moreover, models with a complex acoustic encoder like the conv1d resnet encoder could further reduce PPL. Looking at the CERs, we observed that even with the simple average encoder, a preliminary success could still be obtained. Furthermore, the conv1d resnet encoders reduced the CER by almost 10\%, showing that it is essential to have access to global temporal dependencies before segmentation.

We finally consider the practical decoding scenario. There is a significant performance degradation with the equal segmentation, and it is evidence to the nonlinear relationship of the alignment. Thus, finding an alignment-free approach will be an urgent future work \cite{triggered-attention, cif}. The performance of two conventional ASR systems directly from the original paper of AISHELL-1 \cite{aishell-1} are also listed, and a significant gap exists between our method and the baselines, showing that there is still much room for improvement. Nevertheless, to the best of our knowledge, this is the first study to obtain an ASR system by fine-tuning a pretrained large-scale LM. Moreover, it is worth noticing that the proposed method is readily prepared for n-best re-scoring \cite{bert-asr-rescoring}, though in this paper we mainly focus on building an ASR system. We thus leave this as future work.

\subsection{Error Analysis}

In this section, we examine two possible reasons for the current unsatisfying results.

\subsubsection{Polyphone in Mandarin}

Mandarin is a character-based language, where the same pronunciation can be mapped to different characters. As our method is based on a character-based BERT, it might be infeasible for the model to learn to map the same acoustic signal to different characters. To examine if our model actually suffered from this problem, the syllable error rates (SERs) were calculated and reported in Table~\ref{tab:aishell}. It is obvious that the SERs are much lower than the CERs, showing the existence of this problem. Thus, learning phonetically-aware representations will be a future direction.

\begin{table}[t]
	\centering
	\captionsetup{justification=centering}
	\caption{Development set results with a ratio of the leading characters correctly recognized.}
	\centering
	\begin{tabular}{ c c c c }
		\toprule
		Model & Ratio & CER & SER \\
		\midrule
		\multirow{3}{*}[-1pt]{Conv1d resnet 3} & 0 & 54.8 & 49.6 \\
		& 1/3 & 61.9 & 55.3\\
		& 1/2 & 57.3 & 51.4\\
		\bottomrule
	\end{tabular}
	\label{tab:error-prop}
	\vspace*{-3.5mm}
\end{table}

\subsubsection{Error propagation}

BERT benefits from the self-attention mechanism and is therefore known for its ability to capture global relationship and long-term dependencies. The full-power of BERT may not be exerted with a relatively short context. That is to say, our BERT-ASR can be poor at the earlier decoding steps. As a result, the error in the beginning might propagate due to the recursive decoding process. To examine this problem, we assume that the starting characters up to a certain ratio are correctly recognized, and start the decoding process depending on those characters. Although we expected the error rates to decrease as the ratio increases, as shown in Table~\ref{tab:error-prop}, the CERs and SERs were not lower. Thus, we conclude that error propagation was not a major issue.

\section{Conclusion}

In this work, we proposed a novel approach to ASR by simply fine-tuning BERT, and described the detailed formulation and several essential techniques. 
To verify the proposed BERT-ASR, we demonstrated initial results on the Mandarin AISHELL-1 dataset, and analyzed two possible sources of error. In the future, we will investigate more complex model architectures and the possibility of multi-task learning, in order to close the gap between our and conventional ASR systems. We also plan to evaluate the BERT-ASR on other languages, and apply the proposed method for n-best re-scoring \cite{bert-asr-rescoring}.

\bibliographystyle{IEEEbib}
\bibliography{refs}

\end{document}